\documentstyle[12pt,aasms4,epsfig,rotate]{article}

\newcommand{\mbh}{$M_{\rm BH}$}
\newcommand{\etal}{ {\it et al.}}

\newcommand{\degree}{$^{\circ}$}
\newcommand{\msun}{M_{\odot}}
\newcommand{\mdot}{\dot{M}}
\newcommand{\rg}{r_{\rm g}}

\received{5 February, 1997}
\accepted{8 April, 1997}
\slugcomment{To Appear in The Astrophysical Journal, Letters, Vol. 482, June 20}
\lefthead{Zhang, Cui, and Chen}
\righthead{Black Hole Spin}

\begin{document}

\title{BLACK HOLE SPIN IN X-RAY BINARIES: OBSERVATIONAL CONSEQUENCES}
\author{S. N. Zhang\altaffilmark{1,2}, Wei Cui\altaffilmark{3} and
Wan Chen\altaffilmark{4,5}
}
\altaffiltext{1}{ES-84, NASA/Marshall Space Flight Center, Huntsville,
AL 35812, zhang@ssl.msfc.nasa.gov}

\altaffiltext{2}{Universities Space Research Association}

\altaffiltext{3}{Room 37-571, Center for Space Research, Massachusetts 
Institute of Technology, Cambridge, MA 02139, cui@space.mit.edu}

\altaffiltext{4}{Department of Astronomy, University of Maryland, College
 Park, MD 20742}
\altaffiltext{5}{NASA/Goddard Space Flight Center, Code 661, Greenbelt,
 MD 20771, chen@milkway.gsfc.nasa.gov}

\begin{abstract}

We discuss the observational consequences of black hole spin in X-ray
binaries within the framework of the standard thin accretion disk model.
When compared to theoretical flux distribution from the surface of a thin
disk surrounding a Kerr black hole, the observed X-ray properties of the
Galactic superluminal jet sources, GRO~J1655-40 and GRS~1915+105, strongly
suggest that each contains a black hole spinning rapidly in the same
direction as the accretion disk. We show, however, that some other black
hole binaries with an ultra-soft X-ray component probably harbor only non-
or slowly-spinning black holes, and we argue that those with no detectable
ultra-soft component above 1-2 keV in their high luminosity state may
contain a fast-spinning black hole but with a retrograde disk. Therefore,
all classes of known black hole binaries are united within one scheme.
Furthermore, we explore the possibility that spectral state transitions in
Cyg~X-1 are simply due to temporary disk reversal, which can occur in a
wind accretion system.

\end{abstract}


\section{Introduction}

Significant progress has been made in recent years in observations of
Galactic black hole binaries (BHBs) (cf.\ Tanaka \& Lewin 1995 for a
review).  Optical studies show that the dynamic mass of the compact object
in many black hole candidates (BHCs) is well above 3  $\msun$, the upper
limit on the mass of neutron stars. BHCs are so classified because of the
similarity of their X-ray properties to those of Cyg~X-1, the first
dynamically proven BH with $M>7\msun$. Such similarity includes the
presence of an ultra-soft spectral component with a characteristic
blackbody (BB) temperature of 0.5-2 keV, and an underlying power-law hard
tail that extends beyond 100 keV. Both are rarely seen from NS binaries.

The ultra-soft component is emitted from the inner region of the accretion
disk, close to the BH horizon. The origin of the hard tail is still
unclear, but is believed to be related to the lack of a solid surface on BHs and the
strong gravitational field near the horizon. Systematic optical studies of
BHCs selected based on these criteria have been very successful in finding
compact objects with masses $>3\msun$.

However, an ultra-soft component of 0.5-1.0 keV is absent in the bright
state X-ray spectra of at least 5 (GS~2023+338, GRO~J0422+32, GRO~J1719-24,
1E~1740.7-2942 and GRS~1758-258) of about two dozen known BHCs. Neither is
it observed in Cyg~X-1 and GX339-4 when they are in the so-called hard (or
low) state. Both GS~2023+338 and GRO~J0422+32 are dynamically confirmed
BHBs. It is thus an outstanding puzzle why the ultra-soft component is
missing in their X-ray spectra. None of these sources is an eclipsing
system, so the outer edge of the disk is not likely to block radiation from
the inner disk.

In this letter, we suggest that the strength of the ultra-soft component
is directly related to the BH spin. The presence (or absence) of
such a component would, therefore, depend critically on the specific
angular momentum of the BH and the orientation of its spin axis with
respect to the rotation direction of the accretion disk.

\section{Disk Emission from a Kerr Black Hole Binary}

We assume a geometrically thin, optically thick accretion disk (Novikov \&
Thorne 1973; Shakura \& Sunyaev 1973) orbiting a Kerr BH in the equatorial
plane. The BH spin affects the properties of the inner disk region in
at least two ways (Bardeen, Press, \& Teukolsky 1972). First, the radius of
a Kerr BH horizon, $r_{\rm h}$, is smaller than that of a non-spinning BH,
$r_{\rm h}=\rg+(\rg^{2}-a^{2})^{1/2} = \rg [ 1+ (1-a^2_{\ast})^{1/2}]$,
where $\rg=GM/c^2$ ($M$ is the BH mass), $a=J/Mc$ ($J$ is the BH angular
momentum). The dimensionless specific angular momentum, $a_{\ast} = a/\rg$,
varies from --1 to 1. For maximally spinning BHs, 
$\vert a_{\ast} \vert =1$ and 
$r_{\rm h}(a_{\ast}$=$\pm1)$=$\rg$=0.5$r_{\rm h}(a_{\ast}$=$0)$. Second, the
radius of the last (marginally) stable orbit of the accretion disk is a
function of the BH spin, i.e., 
\begin{equation} 
r_{\rm last} = \rg \{3+A_{2}\pm[(3-A_{1})(3+A_{1}+2A_{2})]^{1/2}\},
\end{equation} 
where
$A_{1}=1+(1-a^{2}_{\ast})^{1/3}[(1+a_{\ast})^{1/3}+(1-a_{\ast})^{1/3}]$,
$A_{2}=(3a^{2}_{\ast}+A_{1}^{2})^{1/2}$; the lower and upper signs are for
a prograde disk (i.e., rotating in the same direction of the BH, or
$a_{\ast}>0$) and a retrograde disk ($a_{\ast}<0$), respectively.
Therefore, a disk around a Kerr BH may extend all the way in to the
horizon, $r_{\rm last}(a_\ast$=1)=$\rg$ or be expelled to $r_{\rm
last}(a_{\ast}$=--1)=9$\rg$, as compared to the canonical Schwarzschild
case, $r_{\rm last}(a_{\ast}$=0)=6$\rg$. 

The local gravitational energy release per unit area from the surface of
the disk is also a function of the BH spin (Page \& Thorne 1974),
\begin{eqnarray}
F(x) & = & \frac{3\mdot c^{6}}{8\pi M^{2} G^{2}}
 \frac{1}{x^{4}(x^{3}-3x+2a_{\ast})}
 \left[x-x_{0} -\frac{3}{2}a_{\ast}\ln (\frac{x}{x_{0}})
 -\frac{3(x_{1}-a_{\ast})^{2}}
{x_{1}(x_{1}-x_{2})(x_{1}-x_{3})}\ln (\frac{x-x_{1}}{x_{0}-x_{1}})
 \right. \nonumber \\
& & \left. -\frac{3(x_{2}-a_{\ast})^{2}}{x_{2}(x_{2}-x_{1})(x_{2}-x_{3})}
\ln (\frac{x-x_{2}}{x_{0}-x_{2}})
 -\frac{3(x_{3}-a_{\ast})^{2}}
{x_{3}(x_{3}-x_{1})(x_{3}-x_{2})}\ln (\frac{x-x_{3}}{x_{0}-x_{3}}) \right],
\end{eqnarray}
where $\mdot$ is the mass accretion rate,
$x=(r/\rg)^{1/2}$, $x_{0}=(r_{\rm last}/\rg)^{1/2}$,
$x_{1}=2\cos(\frac{1}{3}\cos^{-1}a_{\ast}-\pi/3)$,
$x_{2}=2\cos(\frac{1}{3}\cos^{-1}a_{\ast}+\pi/3)$ and
$x_{3}=-2\cos(\frac{1}{3}\cos^{-1}a_{\ast})$.
From eq.~(2), the effective BB temperature of the disk,
$T(x) =(F(x)/\sigma)^{1/4}$ with $\sigma$ being the Stefan-Boltzmann
constant,  actually peaks at an annulus slightly beyond $r_{\rm last}$,
i.e., $r_{\rm peak} = r_{\rm last}/\eta$, and $\eta$ varies slowly from
0.63 to 0.77 for $a_{\ast}$ from --1 to 1. 

The observed disk spectrum, however, bears several important corrections
to this simple formula. First, on the hot inner disk where most X-rays are
emitted, electron scattering may dominate over the free-free absorption
and so causes the color temperature to be greater than the effective
temperature (Ross, Fabian, \& Mineshige 1992). As a result, the inner disk
radiates approximately like a ``diluted'' BB, 
$B(E,f_{\rm col}(x)T(x))/(f_{\rm col}(x))^{4}$ (Ebisuzaki, Hanawa, \&
Sugimoto 1984), where $B(E,f_{\rm col}(x)T(x))$ is the Planck function and
$f_{\rm col}(x)$ is the color correction factor. Detailed calculations
(Shimura \& Takahara 1995), including general relativistic effects, show
that $f_{\rm col}(x)$ depends only very weakly on $x$ so that it can be
approximated by a constant $f_{\rm col}$. Furthermore, $f_{\rm col}$
depends weakly on $M$ and $\mdot$ such that $f_{\rm col} = 1.7 \pm 0.2$ for
$1.4 \le M/\msun \le 10$ and $0.1 \le \mdot/\mdot_{\rm Edd} \le 10$ where
$\mdot_{\rm Edd}$ is the Eddington accretion rate. Therefore, the
dependence of $f_{\rm col}$ on $a_{\ast}$, although unknown, should also be
weak and we adopt  $f_{\rm col}=1.7$ in this paper.

The second important correction is due to general relativistic (GR) effects
near the BH horizon, e.g., the gravitational redshifts and
focusing, which cause both the observed color temperature and integrated
flux to deviate from the local values, depending on the inclination angle
of the disk, $\theta$, and the BH spin (Cunningham 1975). Here we
introduce two additional correction factors, $f_{\rm GR}(\theta,a_\ast)$ --
the fractional change of the color temperature, and $g(\theta,a_\ast)$ --
the additional change of the integrated flux due to viewing angle and GR
effects. We note that in the Newtonian limit $g(\theta,a_\ast) =
\cos(\theta)$, so the pure GR effect on the observed flux is $g_{\rm GR} 
= g(\theta,a_\ast) / \cos(\theta)$. From Cunningham (1975), we derive these
correction factors for several viewing angles and for $a_\ast= 0$ and
0.998. The results are listed in Table 1. The calculations at $a_\ast =
0.998$ are for prograde disks. Retrograde disks should produce less effects
than the $a_\ast = 0$ case. It is clear that the GR effects cause that (1)
the spectrum is redshifted at small $\theta$ but blue-shifted at large
$\theta$, (2) the flux is smaller than the Newtonian flux at small $\theta$
but greater at large $\theta$ (gravitational focusing), and (3) the spin of
the BH induces even larger deviation.

There is a simple relation between the bolometric luminosity of the disk
and the peak emission region,
$L_{\rm disk} \approx 4\pi \sigma r_{\rm peak}^{2} T_{\rm peak}^{4}$
(Makishima \etal\ 1986). We note that $L_{\rm disk}$ depends on both
$\mdot$ and $a_\ast$ but not on $M$. Our numerical integrations over
eq.~(2) confirm that the above relation is accurate to within 10\% for a
wide range of the parameter space. Including the temperature and flux
corrections we introduced above, the observed flux is related to the local
disk luminosity by $F_{\rm earth} = g(\theta,a_\ast) L_{\rm disk} / 2\pi
D^2$, and the observed color temperature is  $T_{\rm col}=f_{\rm
GR}(\theta,a_{\ast})f_{\rm col}T_{\rm peak}$. Therefore,  the inner disk
radius, $r_{\rm last}$, for a source at distance $D$ can be derived as,
\begin{equation}
r_{\rm last}=\eta D \left[ \frac{F_{\rm earth}}
{2 \sigma g(\theta,a_{\ast})} \right]^{1/2}
\left[ \frac{f_{\rm col}f_{\rm GR}(\theta,a_{\ast})}{T_{\rm col}}\right]^2.
\end{equation}
Combining eqs.~(1) and (3), one can thus solve for both $r_{\rm last}$ and
$a_\ast$, if \mbh\ and $\theta$ are known. The largest theoretical
uncertainty in eq.~(3) is from $f_{\rm col}$ for prograde Kerr BHs. Based
on our discussion above, however, the uncertainty in $f_{\rm col}$ is
$\sim10$\% which will cause an error of no more than 20\% in $r_{\rm last}$.

Therefore, the spin of a BH may strongly influence both the disk color
temperature and the disk luminosity, as illustrated in Figs.~1-2 
for typical \mbh\ and $\mdot$. We see that a prograde BHBs shall usually 
have a higher color
temperature than the non-spinning systems, while retrograde BHBs, especially
the more massive ones, will have a softer disk component which may in many
cases escape our detection. On the other hand, the conversion efficiency of
the gravitational energy to radiation, $L_{ \rm disk}/\mdot c^{2}$, is a
function of $a_{\ast}$ {\em only} and increases parabolicly from $\sim$3\%
to $\sim$30\% as $a_{\ast}$ changes from --1 to 1; it is $\sim$6\% for
a Schwarzchild BH (Thorne 1974). To be in line with most X-ray
observations, in Fig.~2 we plot the disk luminosity above 2 keV, 
$L_{\rm D}(>$2 keV) as a function of $a_{\ast}$.

\section{Classification of Black Hole Binaries by Black Hole Spin}

Figs.~1-2 show that the accretion disk spectrum of a BHB becomes
distinctively different only when the BH spins extremely rapidly. It is
thus natural to classify BHBs into three groups, namely, extreme prograde
systems ($a_{\ast}\la 1$), non- or slowly-spinning systems 
($\vert$$a_{\ast}$$\vert$$\simeq$0), and extreme retrograde systems 
($a_{\ast}\ga -1$). 

{\bf Extreme prograde systems.}
This class currently includes GRO~J1655-40 and possibly GRS~1915+105, which are
the only two known Galactic superluminal jet sources. For
GRO~J1655-40, its \mbh\ and $\theta$ have been optically determined as
7.02$\pm$0.22$\msun$  and 69.5\degree$\pm$0\degree.08 (Orosz \& Bailyn
1997). Spectral fitting to its ASCA X-ray spectrum during its 1995 August
outburst gives $kT_{\rm col}$=1.36 keV and an unabsorbed BB flux of
$3.3\times10^{-8}$  ergs s$^{-1}$cm$^{-2}$ (Zhang \etal\ 1997a). Applying
eq.~(3) to a  Schwarzschild BH with $\eta$=0.7, $g$=0.368, and 
$f_{\rm GR}$=1.03, we derive an inner disk radius of 
$r_{\rm last}$=23.3 km or 2.3$\rg$ for a 7$\msun$ BH. We note that
all the observables have very small statistical errors and for a 
Schwarzschild BH the quoted $f_{\rm col}$ and $f_{\rm GR}(\theta,a_\ast)$
are exact; the inferred $r_{\rm last}$ for GRO~J1655-40 thus contains at
most 10\% error. Therefore, it will be extremely difficult to reconcile
the discrepancy between the observed $r_{\rm last}$ and the theoretical
minimum of $6\rg$ for a non-spinning BH. 

The simplest solution is to assume that the BH in GRO~J1655-40 is 
spinning. Solving eq.~(1) and (3) self-consistently with $\eta=0.76$, 
$g = 0.354$ and $f_{\rm GR} = 0.954$, we find that $r_{\rm last}=22.1$
km  and $a_{\ast}=0.93$. Although we have used values of $g$ and
$f_{\rm GR}$ for $a_{\ast} = 0.998$ and $f_{\rm col}=1.7$ in the calculation,
based on our above discussion the lower limit to $a_{\ast}$ is 0.7. 
Therefore we conclude that GRO J1655-40 most probably contains
a Kerr BH spinning at between 70\% to 100\% (most likely value of 93\%)
of the maximum rate.

We note that $a_\ast$ derived above is actually consistent with that
inferred independently from the X-ray timing data. A high frequency QPO of
$\sim$300 Hz was observed in GRO~J1655-40 (Remillard 1997). One of its
possible origins is the trapped $g$-mode oscillations near the inner edge 
of the disk (Okazaki, Kato \& Fukue 1987). By including the GR effects, Perez 
\etal\ (1997)
show that the fundamental frequency of the radial modes is given by 
$f\approx 714(\frac{\msun}{M})F(a_{\ast})$. If the $\sim$300 Hz QPO is
indeed  due to this mode, then $F(a_{\ast})=2.94$, which corresponds to 
$a_{\ast}$=0.95, in agreement with our result. It is, however, not clear
how such models can explain why the QPO is more prominent at higher
energies and why it is only observed when the energy spectrum becomes the
hardest (Remillard 1997).

The case for GRS~1915+105 is less straightforward because its BH mass is 
unknown. However, this source also has a similar BB component and a high 
frequency QPO during its recent outburst. We can estimate the radius of the 
inner edge of the disk using its X-ray spectrum and then, by assuming the 
observed QPO is also due to the fundamental g-mode radial oscillations, 
derive the BH spin self-consistently. Given $T_{\rm col}\simeq$2.27 keV 
and $F_{\rm earth}$ = 4.4$\times10^{-8}$ ergs s$^{-1}$cm$^{-2}$ 
(Belloni \etal\ 1997), the inner disk radius is $\sim$40 km, for a distance
of 12.5 kpc and $\theta$=70\degree\ (Mirabel \& Rodriguez 1994). This 
implies a Schwarzschild BH mass of $\sim4.5\msun$ or an extreme Kerr BH mass
of $\sim$27$\msun$ with a prograde disk. On the other hand, applying the
trapped $g$-mode model to the $\sim$67 Hz QPO (Morgan \etal\ 1996, 1997)
yields $M \sim 11\msun$ for a non-spinning BH and $M \sim 36\msun$ for an
extreme prograde Kerr BH (Nowak \etal\ 1997). The consistency between these two approaches thus
suggests that GRS~1915+105 may contain a Kerr BH of $\sim$30$\msun$ which
is also spinning near the maximum rate. 

{\bf Non- or slowly-spinning systems.}
There are several BHBs whose inner disk radii have been reliably measured 
based on their bright state X-ray spectra (e.g., Tanaka \& Lewin 1995) and 
their $M$\ and $\theta$ have also been determined optically. We use eq.~(1) 
to calculate their $a_\ast$ and the results are listed in Table 2, which 
also includes the two superluminal jet sources. Quite different from 
GRO~J1655-40, these sources show little sign of BH spin. 
%The small
%range of the disk inclination angle (60\degree -70\degree) is probably due
%to selection effect. The reason is that a small $\theta$ would make the 
%ellipsoidal modulations hard to observe, thus preventing a meaningful 
%measurement of both $\theta$ and $M$, while large $\theta$ (thus eclipsing)
%systems are intrinsically rare.
While the theoretical uncertainties in $r_{\rm last}$ for non-spinning BHs
are small, the observational uncertainties in $\theta$ and $M$ for these
systems are greater than those for GRO~J1655-40. It is still intriguing to
see that the BHs in these BHBs {\em appear} to be spinning
slowly. This is consistent with the fact that the three nominal BHBs all
have a lower $T_{\rm col}$ than the two superluminal sources. Fig.~1 shows
that when other system parameters are equal, a prograde disk at $a_\ast
\sim1$ always has a higher $T_{\rm col}$ than the disk at $a_\ast \sim0$
since $r_{\rm last} \propto T_{\rm col}^{-2}$. Thus, although their system
parameters are mostly unknown, most of other ultra-soft BHCs listed in
Table~3.1 of Tanaka \& Lewin (1995) may also be slowly-spinning BHBs.
We do not know, though, if they are slow spinors at birth, or they have
been slowed down, or their spin axis simply lies close to the disk plane.

{\bf Extreme retrograde systems.}
The absence of a detectable ultra-soft component above 1-2 keV in the X-ray
bright state could imply an even softer disk spectrum of $kT_{\rm col} <$
0.3 keV. If so, such a BHB will likely contain an extreme Kerr BH with a
retrograde disk (Fig.~1). While there is yet no proof that they are, we
postulate that GRO~J1719-24, GS~2023+338, and GRO~J0422+32 are such Kerr BH
systems. The existence of the extreme retrograde BHBs can be confirmed
unambiguously only when one detects a weak ultra-soft component at a lower
temperature from those sources in their bright outburst state. A reliable
detection will not be easy, though, because it requires an adequate
detector response down to 0.1 keV and a small interstellar absorption
column.

\section{Spectral State Transitions of Cyg~X-1.}

Cyg~X-1 is a particularly interesting source in our accreting Kerr BH
picture because it shows the characteristics of {\it both} a prograde and a
retrograde BHB. This is mainly related to its distinctive spectral state
transitions, whose nature has not been understood properly. Many models
suggest that the state transitions are accompanied by significant $\mdot$
changes, in conflict with recent  observations which showed approximately
constant total luminosity throughout the state transitions (Zhang \etal\
1997b). Another misconception is that the X-ray spectrum of Cyg~X-1 in its
hard state always contains {\it only} a power law with an energy spectral
index of 0.5-1.0 and a spectral break at 50-100 keV. Recently, however, a
very low temperature BB component was clearly detected by ROSAT
(Balucinska-Church  \etal\  1995) and confirmed by ASCA (Ebisawa \etal\
1996), with {\it kT}$_{\rm col}$$\sim$0.1-0.2 keV and an estimated
luminosity of $\sim 5\times 10^{36}$ erg/s. According to Figs.~1 and 2 for $M$
of $10-20\msun$ and $\mdot$ of $10^{17}-10^{18}$ g s$^{-1}$, the observed
low temperature BB component clearly suggests a retrograde system. On the
other hand, observations in 1996 indicate that the inner disk radius 
dropped by a factor of 2--3 and $\mdot$ increased by less than a factor of
2 when Cyg~X-1 changed from the hard to the soft state 
(Zhang \etal\ 1997b).

The inner disk radius in Cyg~X-1 is likely always at the last stable orbit
since $\mdot$ is always high. If so, we find that the observed decrease of
$r_{\rm last}$ during the hard-to-soft transition can be simply explained
by the {\em reversal} of the disk rotation from a retrograde disk to a
prograde disk, or quantitatively, by $a_{\ast}$ switching from --0.75 to
0.75. A smaller $r_{\rm last}$ also causes an increase of the observed
$T_{\rm col}$ by a factor of $\sim$2.5 and of the observed $L_{\rm disk}$
by a factor of $\la$6 (Figs.~1-2). If the BH were not spinning, the
temperature increase would require a significant increase in $\mdot$ so
that the BB luminosity would have jumped by a factor of 40, which is not
seen in the observations.

\section{Discussion}

In summary, all the observed BHBs can be unified within the framework that
the BH in a binary system may spin up to the maximally allowed speed and
both prograde and retrograde systems exist. The two Galactic superluminal
jet sources are the only known rapidly prograde systems so far, implying
that the formation of the relativistic jets is perhaps related to both the
rapid  spin of the BH and the prograde configuration. The majority of other
known ultra-soft BHBs appear to contain slowly spinning BHs. We have also
suggested that the BHBs without any detectable ultra-soft component above 2
keV may be extreme Kerr BHBs with retrograde disks. 

The canonical BHB Cyg~X-1 may actually switch from a retrograde system in
its normal hard (or low) state to a prograde system in the soft (or high)
state, when the accretion disk reverses its rotation direction temporarily, due
to possibly the unstable nature of the wind accretion from its supergiant
massive companion. Although it may sound absurd at first, the accretion disk reversal
is actually not  at all surprising for Cyg X-1 because it is a wind accreting
system.  Two and three dimensional numerical simulations show that the flip-flop of an
accretion disk rotation can indeed occur in such systems (e.g., Matsuda, Inoue,
\& Sawada 1987; Benensohn, Lamb and Taam 1997; Ruffert 1997). The concept of disk reversal, however, cannot be applied
to low mass BHBs, such as GS1124-683, because their accretion material from
Roche-lobe over-flow gets a strong preference of rotation from the orbital
motion. The state transitions in these systems, often accompanied by
significant $\mdot$ changes, are therefore due to entirely different
mechanisms. Perhaps the lower $\mdot$ causes the inner accretion disk to be
truncated in the quiescent (hard) state and thereby forms advection dominated
accretion flow (e.g., Narayan and Yi 1995). In all our calculations, we
consider only the high $\mdot$ state in which we believe the accretion disk
extends to the last stable orbit.

We have up to this point carefully avoided discussion of hard X-ray 
production in BHBs. It is worth noting, however, that the hard X-ray
luminosity of a prograde system is, on average, much lower than that of a
retrograde system. This seems to imply that the hard X-ray emitting region
is related to the volume between the inner disk boundary and the BH
horizon. For Cyg~X-1 in the soft state, its hard X-ray luminosity ($>$20
keV) is about a factor of 10-20 lower for the observed shrinkage of
the inner disk radius by a factor of 2-3. So, the hard X-ray luminosity
seems to be roughly proportional to the volume inside $r_{\rm last}$. 

We thank G.J. Fishman, B.A. Harmon, J. van Paradijs, K. Ebisawa, N. White and 
R. Remillard for many interesting discussions. 
van Paradijs and White deserve special mention
for carefully reading the manuscript and providing many valuable suggestions.

{\it Note added in proof.} After this paper was submitted, Nick White pointed 
out to us that Shapiro and Lightman (ApJ, 1976, Vol. 204, p.555) had
discussed the possibility that the state transitions of Cyg~X-1 are due 
to reversal of the disk rotation for a marginally stable, wind-fed disk 
around a spinning black hole, an idea originally suggested by J.I. Katz 
(1975) in a private communication. From the observed luminosity change 
and the expected radiation efficiency change due to disk reversal, they 
inferred a black hole spin rate of $a_{\ast} \approx 0.9$, as 
compared with our result of $\sim$0.75 in which we used the most recent 
data and also included the expected BB temperature change.

\newpage

\figcaption[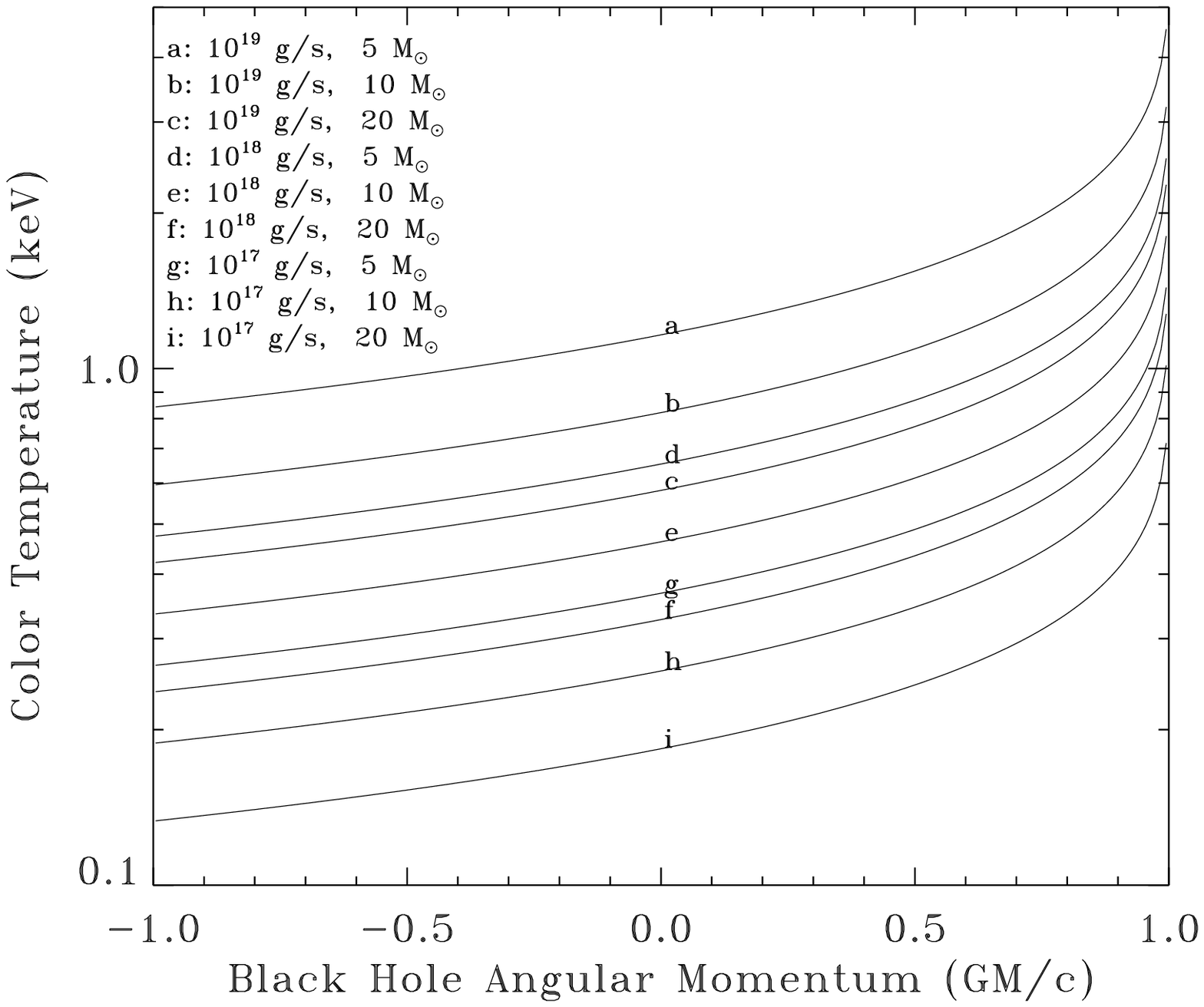]{The peak color temperature ($kT_{\rm col}$) of
the accretion disk emission as a function the dimensionless specific
angular momentum ($a_\ast$) of the Kerr BH, for several mass accretion
rates and BH masses.}

\figcaption[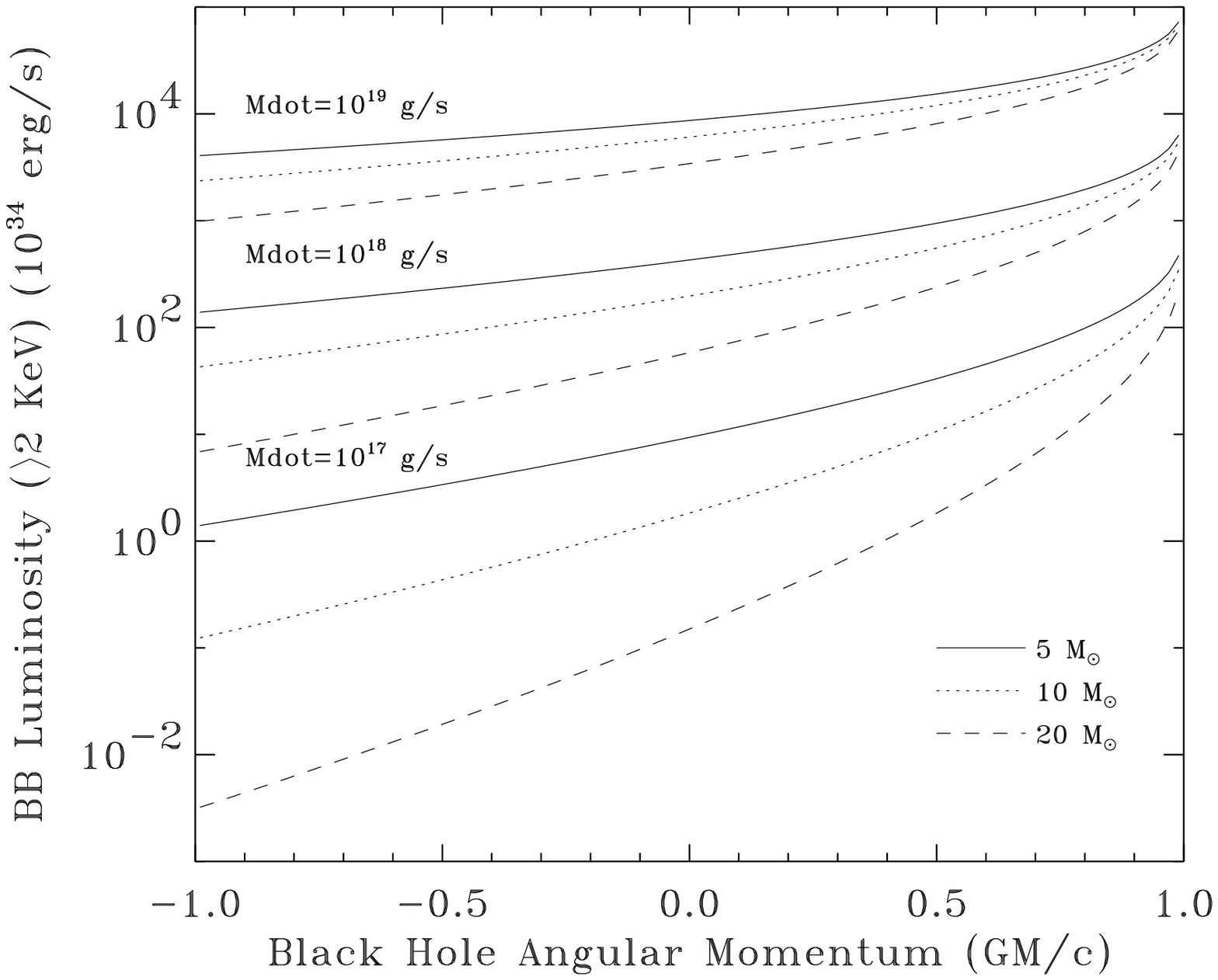]{The ultra-soft component (disk black body)
luminosity above 2 keV, as a function of $a_\ast$, for several mass
accretion rates and black hole masses.}

\newpage

\begin{table}
\begin{tabular} {cccccc}  \hline \hline
 & &\multicolumn{2}{c}{$a_{\ast}$=0.0}
 & \multicolumn{2}{c}{$a_{\ast}$=0.998} \\
 \cline{3-4} \cline{5-6}
$\theta$ & $\cos(\theta)$ & $g(\theta,a_{\ast})$
 & $f_{\rm GR}(\theta,a_{\ast})$ & $g(\theta,a_{\ast})$ 
 & $f_{\rm GR}$($\theta$,$a_{\ast}$) \\ \hline
 0.0 & 1.00 & 0.797 & 0.851 & 0.328 & 0.355 \\
41.4 & 0.75 & 0.654 & 0.870 & 0.344 & 0.587 \\
60.0 & 0.50 & 0.504 & 0.981 & 0.359 & 0.764 \\
75.5 & 0.25 & 0.289 & 1.058 & 0.352 & 1.064 \\
90.0 & 0.00 & 0.036 & 1.354 & 0.206 & 1.657 \\
\hline
\end{tabular}
\caption{General relativistic correction factors to the local
disk spectrum, derived from Cunningham (1975). $a_{\ast}=0.998$
correspondes to the maximally prograde systems. In the Newtonian limit of
$g(\theta,a_{\ast})=\cos(\theta)$ and $f_{\rm GR}(\theta,a_{\ast})=1$.
}
\end{table}

\begin{table}
\begin{tabular} {lcccccc}  \hline \hline
Source & $M$ & $\theta$ & $D$ & $kT_{\rm col}$ & $r_{\rm last}$
 & $a_{\ast}$ \\
 & ($\msun$) & & (kpc) & (keV) & (km) & \\ \hline
1124-68$^{(1)}$ & 6.3 & 60\degree & 2.0 & 1.0 & 57 & -0.04\\
2000+25$^{(2)}$ & 10  & 65\degree & 2.5 & 1.2 & 86 & 0.03\\
LMC X-3$^{(3)}$ & 7   & 60\degree & 50  & 1.2 & 69 & -0.03\\
1655-40$^{(4)}$ & 7   & 70\degree & 3.2 & 1.4 & 22 & 0.93\\
1915+105$^{(5)}$ & $\sim$30 & 70\degree & 12.5 & 2.2 & 40 & $\sim$0.998\\
\hline
\end{tabular}
\caption{The inferred black hole spin in several black hole binaries.
System parameters are from
(1) Orosz \etal\ 1996; (2) Callanan \etal\ 1996; (3) Cowley \etal\ 1983;
(4) Orosz \& Bailyn 1997; (5) Mirabel \& Rodriguez 1994.}
\end{table}

%\end{document}

\newpage

\setcounter{figure}{0}

\begin{figure}
\psfig{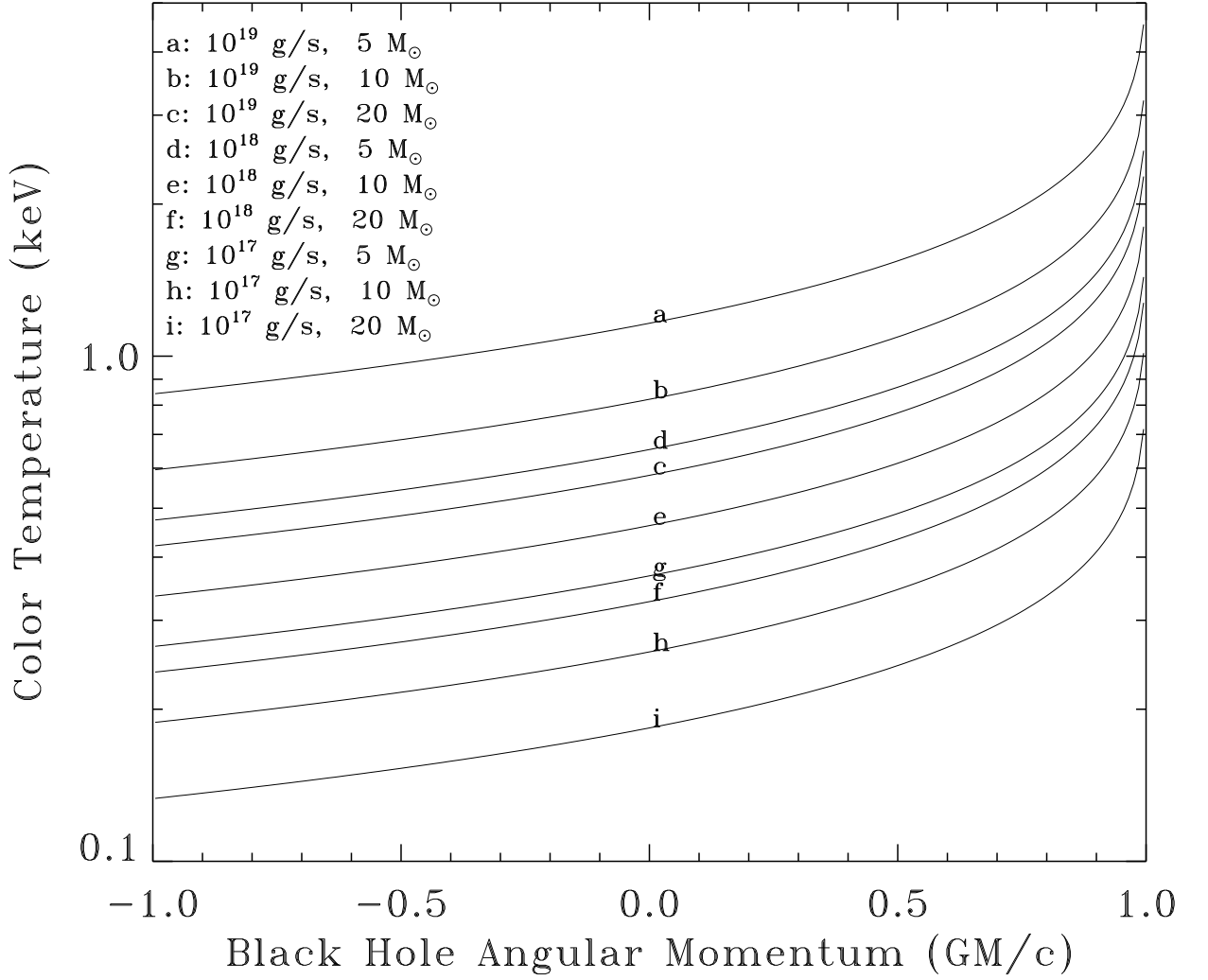}
\caption{The peak color temperature ($kT_{\rm col}$) of
the accretion disk emission as a function the dimensionless specific
angular momentum ($a_\ast$) of the Kerr BH, for several mass accretion
rates and BH masses.}
\end{figure}

\begin{figure}
\psfig{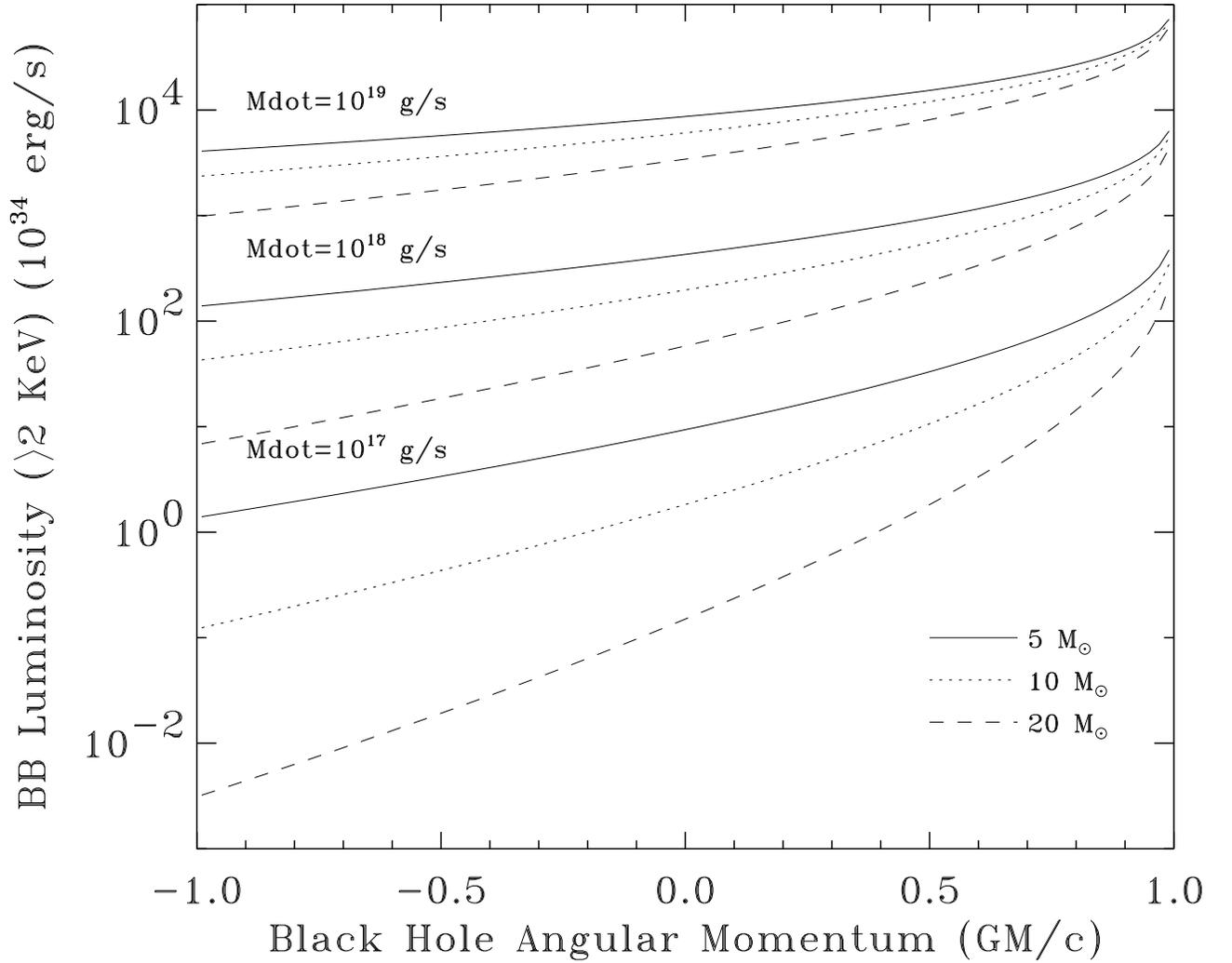}
\caption{The ultra-soft component (disk black body)
luminosity above 2 keV, as a function of $a_\ast$, for several mass
accretion rates and black hole masses.}
\end{figure}

\end{document}